\documentclass{article}

\usepackage{arxiv}

\usepackage[utf8]{inputenc} 
\usepackage[T1]{fontenc}    
\usepackage{hyperref}       
\usepackage{url}            
\usepackage{booktabs}       
\usepackage{amsfonts}       
\usepackage{nicefrac}       
\usepackage{microtype}      
\usepackage{graphicx}
\usepackage{natbib}
\usepackage{doi}
\usepackage{algorithm}
\usepackage{algpseudocode}
\usepackage{todonotes}
\usepackage{fancyhdr}
\usepackage{epstopdf} 
\usepackage{amsmath}
\DeclareMathOperator*{\argmax}{argmax}
\DeclareMathOperator*{\argmin}{argmin}
\usepackage{longtable,tabularx}
\usepackage{multirow}
\usepackage{subcaption}
\usepackage{caption}
\usepackage[]{svg}
\usepackage{setspace}

\title{High-Dimensional Bayesian Optimisation with Large-Scale Constraints via Latent Space Gaussian Processes}

\author{ 
    \href{https://orcid.org/0000-0001-7002-4823}{\includegraphics[scale=0.06]{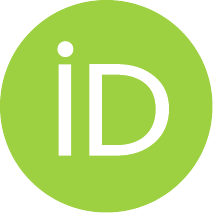}\hspace{1mm}Hauke Maathuis}\thanks{Corresponding author} \\
	Aerospace Engineering\\
	Delft University of Technology\\
	The Netherlands \\
	\texttt{h.f.maathuis@tudelft.nl} \\
	\And
	\href{https://orcid.org/0000-0002-7882-2173}{\includegraphics[scale=0.06]{orcid.pdf}\hspace{1mm}Roeland De Breuker} \\
	Aerospace Engineering\\
	Delft University of Technology\\
    The Netherlands \\
	\texttt{r.debreuker@tudelft.nl} \\
	\And
	\href{https://orcid.org/0000-0001-9711-0991}{\includegraphics[scale=0.06]{orcid.pdf}\hspace{1mm}Saullo G.P. Castro} \\
	Aerospace Engineering\\
	Delft University of Technology\\
    The Netherlands \\
	\texttt{s.g.p.castro@tudelft.nl} \\
}

\renewcommand{\shorttitle}{} 

\begin{document}
\maketitle

\begin{abstract}
        Design optimisation offers the potential to develop lightweight aircraft structures with reduced environmental impact. Due to the high number of design variables and constraints, these challenges are typically addressed using gradient-based optimisation methods to maintain efficiency. However, this approach often results in a local solution, overlooking the global design space. Moreover, gradients are frequently unavailable. Bayesian Optimisation presents a promising alternative, enabling sample-efficient global optimisation through probabilistic surrogate models that do not depend on gradients. Although Bayesian Optimisation has shown its effectiveness for problems with a small number of design variables, it struggles to scale to high-dimensional problems, particularly when incorporating large-scale constraints. This challenge is especially pronounced in aeroelastic tailoring, where directional stiffness properties are integrated into the structural design to manage aeroelastic deformations and enhance both aerodynamic and structural performance. Ensuring the safe operation of the system requires simultaneously addressing constraints from various analysis disciplines, making global design space exploration even more complex. This study seeks to address this issue by employing high-dimensional Bayesian Optimisation combined with a dimensionality reduction technique to tackle the optimisation challenges in aeroelastic tailoring. The proposed approach is validated through experiments on a well-known benchmark case with black-box constraints, as well as its application to the aeroelastic tailoring problem, demonstrating the feasibility of Bayesian Optimisation for high-dimensional problems with large-scale constraints.
\end{abstract}


\section{Introduction}\label{ch:intro}

The design of modern aircraft with enhanced efficiency is crucial for enabling more sustainable aviation. Achieving this involves optimising structural designs to reduce energy consumption. Aeroelastic tailoring emerges as a key technique that has the potential to reduce the weight of aeroelastically efficient high aspect ratio wings. Pioneered by \cite{shirk_aeroelastic_1986}, aeroelastic tailoring incorporates directional stiffness properties to effectively carry and control the aeroelastic deformations. Performing aeroelastic tailoring is a multidisciplinary design and optimisation (MDO) effort, involving aerodynamics for the outer-mould shape definition and calculation of the loads acting over the wing; structural design that usually defines the layout of the main structural components of the wingbox; structural analysis to define and evaluate the relevant failure modes that should be considered as constraints; aeroelasticity that couples the aerodynamic loads with the inertial and elastic properties of the wing in order to characterise the flutter behaviour; and optimisation, to properly explore the design space. Other disciplines are also involved, such as manufacturing, typically resulting in additional constraints for the design variables. 
Evaluating these complex aeroelastic models is computationally expensive, therefore necessitating efficient optimisation algorithms that require fewer analyses before finding an optimum solution. Due to the high number of design variables, describing the structural properties of the system, commonly gradient-based optimisation algorithms are used, leading to an efficient convergence towards the optimal solution. However, the computation of gradients is not always feasible, especially if the model's source code is unavailable. In such cases, the model must be treated as a black box, relying on methods like finite differences to obtain the design sensitivities, which can lead to prohibitively high computational costs that would ultimately motivate the use of gradient-free methods. Furthermore, many engineering problems, such as noisy responses or experimental results, possess inherent complexities that can render gradient-based approaches less effective or even impractical. Additionally, the response surface for feasible designs in aeroelastic tailoring is often multi-modal. This complexity can cause gradient-based methods to become trapped in local optima, overlooking the broader global design space and hindering the discovery of superior designs. Therefore, it is essential to develop methods that efficiently explore the global design space, optimising structures to achieve lighter aircraft configurations. \\~\\
The optimisation problem at hand can be formulated as follows:
\begin{equation}\label{eq:generaloptimisationproblem}
	\min_{\textbf{x} \in \mathcal{X}\subset \mathbb{R}^D} f(\textbf{x}) \ \text{s.t.} \forall i \in \{1,...,G\}, c_i(\textbf{x}) \leq 0, 
\end{equation}
\noindent where $\mathcal{X}\subset \mathbb{R}^D$ is a $D$-dimensional space of potential designs, $f(\textbf{x}): \textbf{x} \in \mathcal{X} \to \mathbb{R}$ the objective function and $G$ constraints arising from the multi-disciplinary analyses. Overall, aeroelastic tailoring can be seen as an optimisation problem consisting of high-dimensional inputs and outputs, where the utilised models are able to map the vector of design variables to the objective function $f(\textbf{x}) \in \mathbb{R}$ and all $G$ constraints $\textbf{c}(\textbf{x}) \in \mathbb{R}^G$. \\~\\ 
The simultaneous consideration of multiple disciplines can lead to large-scale constraints where $G \gg 10^3$, combining buckling, aeroelastic stability, maximum stress, maximum strain, and various others. In aeroelastic tailoring, the optimal stiffness distribution is achieved by means of a sizing optimisation that, in the case of laminated composite wings consists of finding the best set of lamination parameters and the optimum thickness for one or more composite regions \citep{werter_aeroelastic_2017}. Lamination parameters allow a condensed and theoretical representation of the membrane, bending, and coupled stiffness terms of a laminate with continuous variables \citep{dillinger_stiffness_2013}, making the sizing optimisation more convex and more adequate to established continuous optimisation techniques, where the design variables can be treated as continuous variables. Once this sizing optimisation is complete, a second discrete optimisation is performed to retrieve a manufacturable set of ply orientations. Yet, the presence of multiple design regions to maintain design freedom can still result in the number of design variables being in the order of hundreds or thousands. \\~\\
Given the expensive nature of evaluating an aeroelastic model to obtain the objective function values and associated constraints, a sample-efficient optimisation algorithm is crucial. Compared to other gradient-free approaches like Random Search, Genetic Algorithms, and others, Bayesian Optimisation (BO) has proven to be a powerful method for complex and computationally costly problems \citep{mockus_bayesian_1989} and has been extensively applied across various domains, including aircraft design \citep{saves_multidisciplinary_2022}. BO addresses the challenge of expensive evaluations by using computationally inexpensive probabilistic surrogate models, such as Gaussian Processes ($\mathcal{GP}$). These models replace the black-box functions representing the objective and constraints, significantly improving optimisation efficiency \citep{frazier_tutorial_2018}. While many authors have shown that for lower dimensional problems, BO methods perform well, high-dimensional cases pose significant challenges due to the curse of dimensionality \citep{eriksson_high-dimensional_2021,priem_optimisation_2020}, resulting from the fact that high dimensional search spaces are difficult to explore exhaustively. However, BO offers a probabilistic approach to efficiently search the design space to find promising regions and to reduce the computational burden. While these algorithms offer a variety of advantages, including the learning-from-data aspect, uncertainty quantification, the lack of need for gradients, the ability to fuse data in a multi-fidelity context, and the capability to learn the correlation between simulation and experimental data, their scalability to high-dimensional problems with many constraints, as is often the case in engineering design, remains a significant challenge. \\~\\
The present study focuses on employing high-dimensional BO algorithms for aeroelastic tailoring while considering large-scale constraints arising from the multidisciplinary analyses, as formulated in Equation \ref{eq:generaloptimisationproblem}. The novelty of this paper lies in the formulation of a high-dimensional BO method with a dimensionality reduction approach that significantly lowers the computational burden arising from the incorporation of a large number of constraints. Subsequently, the methodology is applied to the $7D$ speed reducer problem with $11$ black box constraints before its application to aeroelastic tailoring is presented. First, Chapter \ref{ch:hdbo} introduces $\mathcal{GP}$s as a surrogate modelling technique, as well as BO for unconstrained and constrained problems. Additionally, the difficulties in terms of scalability are highlighted. Subsequently, dimensionality reduction in the context of constrained BO is presented in Chapter \ref{ch:large-scale-constraints}, before the theory is applied to the aeroelastic tailoring optimisation problem in Chapter \ref{ch:application}. \\~\\

\section{High-Dimensional Constrained Bayesian Optimisation}\label{ch:hdbo}
This section briefly introduces BO within the context of high dimensionality and constraints. $\mathcal{GP}$s are introduced as the herein employed surrogate modelling technique. Subsequently, $\mathcal{GP}$s are linked to unconstrained BO, which is then expanded to address the constrained scenario, followed by an outline of the challenges encountered in this work.

\subsection{Gaussian Processes}\label{ch:gp}
A $\mathcal{GP}$ in the context of BO serves as a probabilistic surrogate model that efficiently represents an unknown function $f(\textbf{x})$. Recall that $\mathcal{X} \subset \mathbb{R}^D$ is a $D$-dimensional domain and the corresponding minimisation problem is presented in Equation \ref{eq:generaloptimisationproblem}. Beginning with a Design of Experiments (DoE) denoted by $\mathcal{D}_0 = \{ \textbf{x}_i,f(\textbf{x}_i)\}_{i=1,...,N}$, where $\textbf{x}_i \in \mathcal{X} \subset \mathbb{R}^D$ is the $i$-th of $N$ samples and $f (\textbf{x}_i):\mathcal{X} \to \mathbb{R}$ the objective function, mapping from the design space to a scalar value. $\mathcal{GP}$s are commonly employed within BO to construct a surrogate model $\hat{f}(\textbf{x}): \mathcal{X} \to \mathbb{R}$ of the objective function $f$ from this given data set $\mathcal{D}$. Therefore, it is assumed that the objective function $f$ follows a $\mathcal{GP}$, which is also called a multivariate normal distribution $\mathcal{N}$. By defining the mean $m:\mathcal{X} \to \mathbb{R}$ and covariance  $k:\mathcal{X} \times \mathcal{X} \to \mathbb{R}$, a noise-free surrogate can thus be denoted as:
\begin{equation}\label{eq:prior}
	f(\textbf{x}) \sim \hat{f}(\textbf{x}) \mid \mathcal{D} = \mathcal{GP} \left( m( \textbf{x}) , k(\textbf{x}, \textbf{x}') \right),
\end{equation}
also known as the prior. The prior encapsulates the initial belief that observations are normally distributed. A common choice for the covariance function, also called kernel, is the squared exponential kernel $k(\textbf{x},\textbf{x}')$ defined as 
\begin{equation}\label{eq:kernel}
	k(\textbf{x},\textbf{x}') = s^2 \exp{\left( -\frac{1}{2} \sum_{i=1}^{D} \left( \frac{x_i-x_i'}{l_i} \right)^2 \right)},
\end{equation}
\noindent which encodes the similarity between two chosen points $\textbf{x}$ and $\textbf{x}'$ \cite{rasmussen_gaussian_2006}. The parameter $l_i$ for $i=1,...,D$ is called the length scale and measures the distance for being correlated along $x_i$. Together with $\sigma^2$, often called the signal variance, the parameters form a set of so-called hyperparameters $\boldsymbol{\theta} = \{ l_1,...,l_D,\sigma^2\}$, in total $D+1$ parameters, which need to be determined to train the model with respect to the target function. The kernel matrix is defined as $\textbf{K} = \left[ k(\textbf{x}_i,\textbf{x}_j)  \right]_{i,j=1,...,N} \in \mathbb{R}^{N \times N}$. The kernel needs to be defined such that $\textbf{K}$ is symmetric positive definite to ensure its invertibility. The positive definite symmetry is guaranteed if and only if the used kernel is a positive definite function, as detailed in \cite{schoenberg_metric_1938}. 
The values of the hyperparameters $\boldsymbol{\theta}$ are determined by maximising the marginal likelihood, written as 
\begin{equation}
	\log p(\mathbf{f} \mid \mathcal{D},\boldsymbol{\theta}) = -\frac{1}{2} \mathbf{f}^{\top} \textbf{K}^{-1} \mathbf{f}-\frac{1}{2} \log \left| \textbf{K} \right|-\frac{n}{2} \log 2 \pi.
\end{equation}
Computing the partial derivative with respect to the hyperparameters $\boldsymbol{\theta}$ gives
\begin{equation}
	\frac{\partial}{\partial \theta_j} \log p(\mathbf{f} \mid \mathcal{D},\boldsymbol{\theta}) = \frac{1}{2} \mathbf{f}^{\top} \textbf{K}^{-1} \frac{\partial \textbf{K}}{\partial \theta_j} \textbf{K}^{-1} \mathbf{f}-\frac{1}{2} \operatorname{tr}\left(\textbf{K}^{-1} \frac{\partial \textbf{K}}{\partial \theta_j}\right) 
\end{equation}
\noindent which can be used within a gradient-based optimisation for model selection or in other words, hyper-parameter tuning. More detailed information can be found in \cite{rasmussen_gaussian_2006}.\\~\\ 
Considering a new query point $\textbf{x}_{+} \in \mathcal{X}$, the stochastic process in Equation \ref{eq:prior} can be used to predict the new query point 
\begin{equation}
	f(\mathbf{x}_+)\mid \mathcal{D} \sim \mathcal{N} \left( \mu(\textbf{x}_{+}),k(\textbf{x}_{+},\textbf{x}_{+}) \right).
\end{equation}
The posterior mean $\mu(\bullet)$ and covariance function $\sigma(\bullet)$ are computed by
\begin{align}\label{eq:posterior} 
	\mu(\textbf{x}_{+}) &= \textbf{k}(\textbf{x}_{+},\textbf{X}) \textbf{K}(\textbf{X},\textbf{X})^{-1} \textbf{f}, \\
	\sigma(\textbf{x}_{+}) &=  k(\textbf{x}_{+},\textbf{x}_{+})-\textbf{k}(\textbf{x}_{+},\textbf{X}) \textbf{K}(\textbf{X},\textbf{X})^{-1} \textbf{k}(\textbf{X},\textbf{x}_{+}),
\end{align}
\noindent where $\textbf{X} = [ \textbf{x}_1,\textbf{x}_2,...,\textbf{x}_N] \subset \mathcal{D}$ is the collection of samples and $\textbf{f} = [f_1,f_2,...,f_N] \subset \mathcal{D}$ of computed objective values in $\mathcal{D}$. 

\subsection{Unconstrained Bayesian Optimisation}\label{ch:unconstr_bo}
Up to this stage, the $\mathcal{GP}$ has been computed using the initial samples contained in $\mathcal{D}_0$. BO now proceeds iteratively to enhance the accuracy of the surrogate model by enriching $\mathcal{D}$ while exploring the design space. Thus, leveraging the acquired data, the endeavour is to identify regions expected to yield optimal values. The problem at hand can be written as 
\begin{equation}
	\min_{\textbf{x} \in \mathcal{X}} f(\textbf{x}).
\end{equation}
An acquisition function $\alpha: \mathcal{X} \to \mathbb{R}$ is used to guide the optimisation through the design space while trading off exploration and exploitation based on the posterior mean and variance defined in Equation \ref{eq:posterior}. The former describes the exploration of the whole design space, whereas the latter tries converging to an optimum based on the data observed. This can be written as 
\begin{align}\label{eq:acquisition}
    \textbf{x}_{+} \in \argmax_{\textbf{x} \in \mathcal{X}} \alpha (\textbf{x} \mid \mathcal{D}).
\end{align}
Numerous acquisition functions exist, often making use of the predictive mean $\hat{\mu}(\mathbf{x})$ and variance $\hat{\sigma}(\mathbf{x})$. Popular choices for such an acquisition function are for example Expected Improvement (EI) \citep{mockus_j_application_1978} or Thompson Sampling (TS) \citep{thompson_likelihood_1933}.  


\subsection{Constrained Bayesian Optimisation}\label{ch:constr_bo}
Most engineering design problems involve constraints, which can be integrated into the previously introduced BO method, discussed in e.g. \cite{gardner_bayesian_2014,gelbart_bayesian_2014,hernandez-lobato_general_2016}. Assuming that the output of a model evaluation at design point $\textbf{x}_i$ includes not only the objective function $f(\textbf{x}_i)$, but also a mapping from the design space to a collection of $G$ constraints $\textbf{c}(\textbf{x}_i): \mathcal{X} \to \mathbb{R}^G$. Consequently, the DoE for this scenario is represented as $\mathcal{D} = \{ \textbf{x}_i, f(\textbf{x}_i), \textbf{c}(\textbf{x}_i) \}_{i=1,..,N}$. The new design point found needs to lie in the feasible space $\mathcal{X}_f$, written as $\textbf{x}_{+} \in \mathcal{X}_f \subset \mathcal{X}$ where $\mathcal{X}_f := \{ \textbf{x} \in \mathcal{X} \mid \hat{c}_j(\textbf{x}) \leq 0, \ j=1,...,G \}$.
\cite{gardner_bayesian_2014} propose modelling each constraint $c_j(\textbf{x}), j=1,...,G$ with an independent surrogate model, akin to how the objective function is modelled:
\begin{equation}\label{eq:prior_c}
	c_j(\textbf{x}) \sim \hat{c}_j(\textbf{x}) \mid \mathcal{D} = \mathcal{GP} \left( m( \textbf{x}) , k(\textbf{x}, \textbf{x}') \right) = \mathcal{N} \left( \mu(\textbf{x}),\sigma( \textbf{x})^2 \right),
\end{equation}
\noindent leading to $G+1$ $\mathcal{GP}$ models in total, enabling the extension of Equation \ref{eq:acquisition} for constrained problems, written as
\begin{equation}
	\alpha_c(\mathbf{x} \mid \mathcal{D}) = \alpha(\mathbf{x} \mid \mathcal{D}) \prod_{j=1}^{G} \mathbb{P} \left( \hat{c}_j(\textbf{x} \leq 0 )\right),
\end{equation}
involving the probability $\mathbb{P}$ that the $j$-th constraint is not violated. Accordingly, within the acquisition strategy, the sub-problem
\begin{equation}\label{ch:newpoint}
	\textbf{x}_{+} \in \argmax_{\textbf{x} \in \mathcal{X}_f \subset \mathcal{X}} \alpha_c(\mathbf{x} \mid \mathcal{D})
\end{equation}
has to be solved. This subsection serves to introduce the fundamental aspects of constrained BO concisely, emphasising that each constraint must be modelled via a separate $\mathcal{GP}$ model. Of course, a multitude of constrained acquisition functions exist. Among these approaches, for instance, is the use of Thompson Sampling as an acquisition function \citep{hernandez-lobato_parallel_2017}, extended to the constrained setting in \cite{eriksson_scalable_2020}. A major advantage is its scalability to larger batch sizes. The latter study also demonstrates the superiority of this approach compared to EI which is why constrained TS is employed in the course of this work and is explained in Algorithm \ref{alg:ts}. Therein, for each $\mathcal{GP}$ used for modelling the objective function and the $G$ constraints, the posterior is computed. For a batch size of $Q$ points, a sample is drawn to get a realisation of the surrogate models. Then, $N_c$ candidate points are evaluated on the $\mathcal{GP}$s to obtain either a set of feasible points with optimal objective value or points with a minimum total constraint violation $\mathbf{X}_+$. 

\begin{algorithm}
	\begin{algorithmic}
		\caption{Constrained Thompson Sampling}\label{alg:ts}
		\State \textbf{Input:} $\mathcal{D}_k$ of $k$-th iteration, $Q$ batch size,  $\textbf{X}_{c} = [ \textbf{x}_1, \textbf{x}_2,  ... , \textbf{x}_{N_c} ]$ with $N_c$ candidates 
		\While{Computational budget is not exhausted}
        \State $\mathbf{X}_+=\{\}$
		\State Compute current posterior p($\boldsymbol{\theta} | \mathcal{D}_k$) for $f,c_1,...,c_G$
		\For{q = 1:Q}
		\State Sample $\boldsymbol{\theta}$ from p($\boldsymbol{\theta} | \mathcal{D}_k)$ to obtain realisations for $\hat{f},\hat{c}_1,...,\hat{c}_G$
		\State Evaluate $\{\textbf{x}_i \mid i \in \mathbb{N},\;1 \leq i \leq N_c\}$ on $\hat{f}(\textbf{x}_i), \hat{c}_1(\textbf{x}_i),...,\hat{c}(\textbf{x}_i)$ from the respective posterior distribution
		\State Obtain $\hat{f}(\textbf{x}_i), \hat{c}_1(\textbf{x}_i),...,\hat{c}_G(\textbf{x}_{i})$
		\State Choose $\mathbb{X}_f = \{ \textbf{x}_i | \hat{c}_l(\textbf{x}_i) \leq 0 \ \text{for} \ 1 \leq l \leq G \}$
		\If{$\mathbb{X}_f \neq \emptyset$}
		$\textbf{x}_+^{q} = \argmax_{\textbf{x} \in \mathbb{X}_f} \hat{f}(\textbf{x})$
		\Else \ Obtain the minimum of total violation by computing $\textbf{x}_+^{q} = \argmin_{\textbf{x} \in \textbf{X}_c} \sum_{i=1:G} \max(\hat{c}_i(\textbf{x}),0)$
		\EndIf
		\State $\textbf{X}_+ = \textbf{X}_+ \cup \{\textbf{x}_+^{q}\}$
        \EndFor
		\EndWhile
	\end{algorithmic}
\end{algorithm}

\subsection{High-Dimensional Bayesian Optimisation: Challenges and Advances} 
BO algorithms consist of two main components, namely the probabilistic surrogate model, $\mathcal{GP}$s, which are based on Bayesian statistics \citep{rasmussen_gaussian_2006}, and an acquisition function to guide the selection where to query the next point to converge towards the minimiser of the objective function. While these algorithms have been proven to be very efficient for lower-dimensional problems \citep{binois_survey_2022}, scaling them to higher dimensions implies some difficulties:
\begin{itemize} 
	\item The curse of dimensionality dictates that as the number of dimensions increases, the size of the design space grows exponentially, making an exhaustive search impractical.
	\item With higher dimensions, there is an increase in the number of tunable hyperparameters \(\boldsymbol{\theta} \in \mathbb{R}^{D+1}\), resulting in a more cumbersome \(\mathcal{GP}\) model learning, possibly leading to increased uncertainty.
	\item Higher-dimensional problems necessitate more samples \(N\) to construct an accurate surrogate model. The inversion of the covariance matrix \(\textbf{K} \in \mathbb{R}^{N \times N}\) becomes computationally intensive with a complexity for inference and learning of \(\mathcal{O}(N^3)\) and \(\mathcal{O}(N^2)\) for memory.
	\item Insufficient data collection results in sparse sampling across the \(D\)-dimensional hyperspace, causing samples to be widely dispersed from each other. This dispersion hinders effective correlation among the samples.
	\item Acquisition function optimisation faces increased uncertainty in high-dimensional settings, requiring more evaluations of the surrogate model \citep{binois_survey_2022}.
\end{itemize} 
Various strategies have been employed to address the challenge of high-dimensional input spaces in scenarios with few or no constraints. In \cite{wang_bayesian_2016}, random projections are utilised to reduce high-dimensional inputs to a lower-dimensional subspace, allowing for the construction of the $\mathcal{GP}$ model directly in this reduced space, thereby reducing the number of hyperparameters. Similarly, \cite{raponi_2020, antonov_2022} employ (kernel) Principal Component Analysis on the input space to identify a reduced set of dimensions based on evaluated samples, followed by training the surrogate model in this reduced dimensional space. In contrast, \cite{eriksson_high-dimensional_2021} adopt a hierarchical Bayesian model that assumes varying importance among design variables, using a sparse axis-aligned prior on the length scale to discard dimensions unless supported by accumulated data. However, \cite{santoni_2023} demonstrates high computational overhead in this approach. Additionally, decomposition techniques, such as additive methods, are employed to partition the original space, as demonstrated in \cite{kandasamy_high_2016, ziomek_are_2023}. \\~\\
The Trust-Region Bayesian Optimisation (TuRBO) algorithm, described in \cite{eriksson_scalable_2020}, takes a different route where the design space is partitioned into multiple independent trust regions. Results from \cite{eriksson_scalable_2020} demonstrate promising outcomes for this approach, particularly in high-dimensional problems where gathering sufficient data to construct a globally accurate surrogate model is challenging due to the curse of dimensionality. Instead, surrogates are focused on these defined trust regions, which adjust in size during optimisation. Trust regions are defined as hyper-rectangles of size \( L \in \mathbb{R} \), centred at the best solution found so far and initialised with \( L \leftarrow L_{init} \), a user-defined parameter. The size \( L_{TR} \) of each trust region is determined using the length scale \( l_i \) of the $\mathcal{GP}$, defined in Equation \ref{eq:kernel}, and a base length scale \( L \): 
\begin{equation}\label{eq:TRlength}
	L_{TR} = \frac{l_i L}{\left( \prod_{j=1}^{D} l_j \right)^{1/D}}.
\end{equation}
In each optimisation iteration, a batch of \( q \) samples are drawn within the trust region (TR). When the design space is normalised to \( \mathcal{X} \in [-1,1] \) and \( L \) spans the entire design space with \( L \rightarrow 2 \) kept constant, the Trust Region approach resembles a standard BO algorithm as outlined in \cite{frazier_tutorial_2018}. The evolution of \( L \) significantly influences the convergence of this method, and specific hyperparameters governing its adaptation are detailed in \cite{eriksson_scalable_2020}.\\~\\
All the algorithms previously discussed focus exclusively on unconstrained optimization problems. The Trust Region approach, however, addresses constraints explicitly by adapting the batched Thompson Sampling method from \cite{thompson_likelihood_1933} as an acquisition function for constrained problems \cite{eriksson_scalable_2021}, detailed in Algorithm \ref{alg:ts}. This method, known as Scalable Constrained Bayesian Optimisation (SCBO), employs separate \(\mathcal{GP}\)s to model each constraint within the current Trust Region. Scaling BO to high-dimensional problems necessitates addressing significant challenges through specific assumptions. While existing approaches demonstrate promising results, handling large-scale constraints, such as those encountered in aircraft design problems where \( G > 10^3 \), remains insufficiently addressed. This work adopts the constrained TuRBO algorithm SCBO for high-dimensional BO due to its explicit treatment of constraints. Next, an extension of this method is introduced to address the challenge posed by large-scale constraints.

\section{Large-Scale Constrained Bayesian Optimisation via Latent Space Gaussian Processes}\label{ch:large-scale-constraints}
Recall the optimisation problem formulated in Equation \ref{eq:generaloptimisationproblem}. By using constrained BO methods, as shown earlier, each of the $G$ constraints needs to be modelled with an independent $\mathcal{GP}$, denoted as $\hat{c}_i(\textbf{x})$. This work follows the idea of \cite{higdon_computer_2008} to construct the surrogates on a lower dimensional, latent output space. Let \( \mathcal{V} \subset \mathbb{R}^G \) denote a \( G \)-dimensional space. The objective of this work is to identify a latent space \( \mathcal{V}' \subset \mathbb{R}^g \) such that \( \mathcal{V}' \subset \mathcal{V} \), where \( g \ll G \). This subspace may be found by using dimensionality reduction methods such as Principal Component Analysis (PCA) \citep{jolliffe_principal_2016} on the training data in $\mathcal{D}_k$. An extended nonlinear version of PCA is the kernel PCA (kPCA), presented by \cite{scholkopf_nonlinear_1998}. \\~\\
During the DoE, alongside the samples \(\textbf{x}_i\) and their corresponding objective function values \(f_i\), constraint values \(\textbf{c}:\mathcal{X} \to \mathbb{R}^G\) are also collected in \(\mathcal{D}\). This enables the construction of a matrix \(\textbf{C}(\textbf{x})\) given by:
\begin{equation}
	\textbf{C}(\textbf{x})= 
	\begin{bmatrix} \textbf{c}(\textbf{x}_1)^{\top} \\ \textbf{c}(\textbf{x}_2)^{\top} \\ \vdots \\ \textbf{c}(\textbf{x}_N)^{\top} \end{bmatrix} =
	\begin{bmatrix} c_1(\textbf{x}_1) & c_2(\textbf{x}_1) & ... & c_G(\textbf{x}_1) \\ 
		c_1(\textbf{x}_2) & c_2(\textbf{x}_2) & ... & c_G(\textbf{x}_2) \\
		\vdots & \vdots & \ddots & \vdots \\
		c_1(\textbf{x}_N) & c_2(\textbf{x}_N) & ... & c_G(\textbf{x}_N) \end{bmatrix} 
	\in \mathbb{R}^{N \times G}.
\end{equation}
Here, \(N\) represents the number of samples and \(G\) denotes the number of constraints.

\subsection{Principle Component Analysis (PCA)}
Within PCA, a linear combination with maximum variance is sought, such that 
\begin{equation}
	\textbf{C} \textbf{v} = \lambda \textbf{v}
\end{equation}
\noindent where $\textbf{v}$ is a vector of constants. These linear combinations are called the principle components of the data contained in $\textbf{C}$. After centering the data with $\bar{\textbf{C}} = \textbf{C} - \textbf{I}_N \mu$ with $\mu = \frac{1}{N} \sum_{i=1}^{N} \textbf{c}_i$, a covariance matrix $\mathcal{C}$ is computed
\begin{equation}
    \mathcal{C} = \frac{1}{N-1} \bar{\textbf{C}}^\top \bar{\textbf{C}} \in \mathbb{R}^{G \times G}.
\end{equation}
Subsequently, PCA seeks the set of orthogonal vectors that capture the maximum variance in the data. This is achieved by performing an eigenvalue decomposition of $\mathcal{C}$, to obtain the corresponding eigenvalues $\lambda$ and eigenvectors $\textbf{v}$ such that
\begin{equation}
    \mathcal{C}\textbf{v}_i = \lambda_i \textbf{v}_i, \ \forall i=1,2,...,G
\end{equation}
with $\lambda_1 \geq \lambda_2 \geq ... \geq \lambda_G \geq 0$. The eigendecomposition of $\mathcal{C}$ is then written as 
\begin{equation}
    \mathcal{C} = \boldsymbol{\Psi}\boldsymbol{\Lambda}\boldsymbol{\Psi}^{-1}
\end{equation}
The matrix $\boldsymbol{\Psi} = [ \Psi_1,...,\Psi_G ]\in \mathbb{R}^{G \times G}$ has orthonormal columns such that $\boldsymbol{\Psi}^{\top}\boldsymbol{\Psi} = \boldsymbol{I}_{\boldsymbol{\Psi}}$ and $\boldsymbol{\Lambda} = diag(\lambda_1,...,\lambda_G) \in \mathbb{R}^{G \times G}$ is a diagonal matrix, containing the eigenvalues. By investigating the eigenvalues in $\boldsymbol{\Lambda}$, and choosing the ones with the $g$ highest values, the truncated decomposition is obtained, consisting of the reduced basis containing $g$ orthogonal basis vectors in $\boldsymbol{\Psi}_g \in \mathbb{R}^{G \times g}$ with $g \ll G$.  The new basis vectors can subsequently be used as a projection $\boldsymbol{\Psi}_g: \mathcal{V} \subset \mathbb{R}^G \to \mathcal{V}' \subset \mathbb{R}^g$ to project the matrix $\textbf{C}$ onto the reduced subspace $\tilde{\textbf{C}} \in \mathbb{R}^{N \times g}$, written as 
\begin{equation}\label{eq:PCAprojection1}
	\tilde{\textbf{C}} = \textbf{C} \boldsymbol{\Psi}_g.
\end{equation}
Summarising, the $G$ constraints $\textbf{c}(\textbf{x})$ can be represented on a reduced subspace through the mapping $\boldsymbol{\Psi}_g$ while the eigenvalues $\lambda_i$ give an indication about the loss of information, potentially drastically lowering the number of constraints that need to be modelled. A graphical interpretation is depicted in Figure \ref{fig:pca_graphical}. For a more thorough derivation of this method, the reader is referred to \cite{jolliffe_principal_2016}.

\subsection{Kernel Principle Component Analysis (kPCA)}
While PCA can be seen as a linear dimensionality reduction technique, in \cite{scholkopf_nonlinear_1998} the authors present an extension, called kernel PCA, using a nonlinear projection step to depict nonlinearities in the data. Similarly to the PCA algorithm, the starting point is the (centred) samples $\textbf{c}_i(\textbf{x}_i) \in \mathcal{V} \subset \mathbb{R}^G \ \forall i \in \{ 1,...,N \}$.\\
Let $\mathcal{F}$ be a dot product space (in the following, also called feature space) of arbitrary large dimensionality. A nonlinear map $\boldsymbol{\phi}(\textbf{x})$ is defined as $\boldsymbol{\phi}: \mathbb{R}^G \to \mathcal{F}$.  This map is used to construct a covariance matrix $\mathcal{C}$, similar to PCA, defined as 
\begin{equation}
	\mathcal{C} = \frac{1}{N} \sum_{i=1}^{N} \boldsymbol{\phi}(\textbf{c}(\textbf{x}_i)) \boldsymbol{\phi}(\textbf{c}(\textbf{x}_i))^{\top}.
\end{equation}
The corresponding eigenvalues and eigenvectors in $\mathcal{F}$ are computed by solving
\begin{equation}\label{eq:eig-in-feature-space}
	\mathcal{C}\textbf{v} = \lambda \textbf{v}.
\end{equation}
As stated earlier, since the function $\boldsymbol{\phi}$ maps possibly to a very high-dimensional space $\mathcal{F}$, solving the eigenvalue problem therein may be costly. A workaround is used to avoid computations in $\mathcal{F}$. Therefore, similar to the formulation of the $\mathcal{GP}$ models in Section \ref{ch:gp}, a kernel $k: \mathbb{R}^G \times \mathbb{R}^G \to \mathbb{R}$ is defined as
\begin{equation}\label{eq:pcakernel}
	k(\textbf{c}(\textbf{x}_i),\textbf{c}(\textbf{x}_j)) = \bigl \langle \boldsymbol{\phi}(\textbf{c}(\textbf{x}_i)), \boldsymbol{\phi}(\textbf{c}(\textbf{x}_j)) \bigr \rangle =  \boldsymbol{\phi}(\textbf{c}(\textbf{x}_i))^{\top}\boldsymbol{\phi}(\textbf{c}(\textbf{x}_j))
\end{equation}
and the corresponding kernel matrix $\textbf{K}_{ij}$ as
\begin{equation}
	\textbf{K}_{ij} := \left( \boldsymbol{\phi}(\textbf{c}(\textbf{x}_j)), \boldsymbol{\phi}(\textbf{c}(\textbf{x}_j)) \right) \in \mathbb{R}^{N \times N}.
\end{equation}
By solving the eigenvalue problem for non-zero eigenvalues
\begin{equation}
	\textbf{K} \boldsymbol{\alpha}_i = \lambda_i \boldsymbol{\alpha}_i
\end{equation}
the eigenvalues $\lambda_1 \geq \lambda_2 \geq ... \geq \lambda_N$ and eigenvectors $\boldsymbol{\alpha}^1, ..., \boldsymbol{\alpha}^N$ are obtained. This part can be seen as the linear PCA, as presented before, although in the space $\mathcal{F}$. To map a test point $\textbf{c}_+(\textbf{x})$ from the feature space $\mathcal{F}$ to the $q$-th principle component $\textbf{v}^q$ of Equation \ref{eq:eig-in-feature-space}, the following relationship is evaluated
\begin{equation}\label{eq:KPCAprojection}
	\left( (\textbf{v}^q)^{\top} \boldsymbol{\phi}(\textbf{c}_+(\textbf{x})) \right) = \sum_{i=1}^{N} \boldsymbol{\alpha}_{i}^{q}( \boldsymbol{\phi}(\textbf{c}(\textbf{x}_i))^{\top}  \boldsymbol{\phi}(\textbf{c}_+(\textbf{x}))) \equiv \tilde{\textbf{c}}_+(\textbf{x}_+).
\end{equation}
A graphical interpretation can be found in Figure \ref{fig:kpca_graphical}. The kernel function in Equation \ref{eq:pcakernel} can also be replaced by another a priori chosen kernel function.

\subsection{Dimensionality Reduction for Large-Scale Constraints}
\begin{figure}[h]
	\centering
	\begin{subfigure}{.30\textwidth}		\centering
		\includegraphics[width=\textwidth]{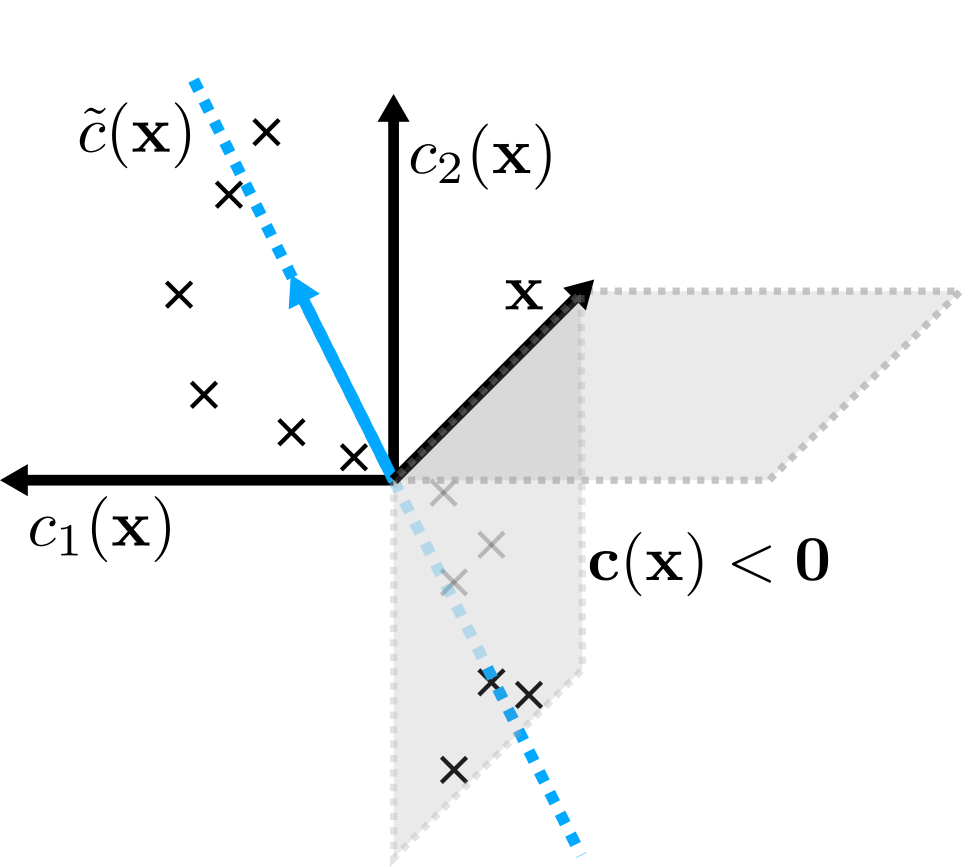}
		\caption{Principal Component Analysis}
		\label{fig:pca_graphical}
	\end{subfigure}
	\hfill
	\begin{subfigure}{.60\textwidth}
		\centering
		\includegraphics[width=\textwidth]{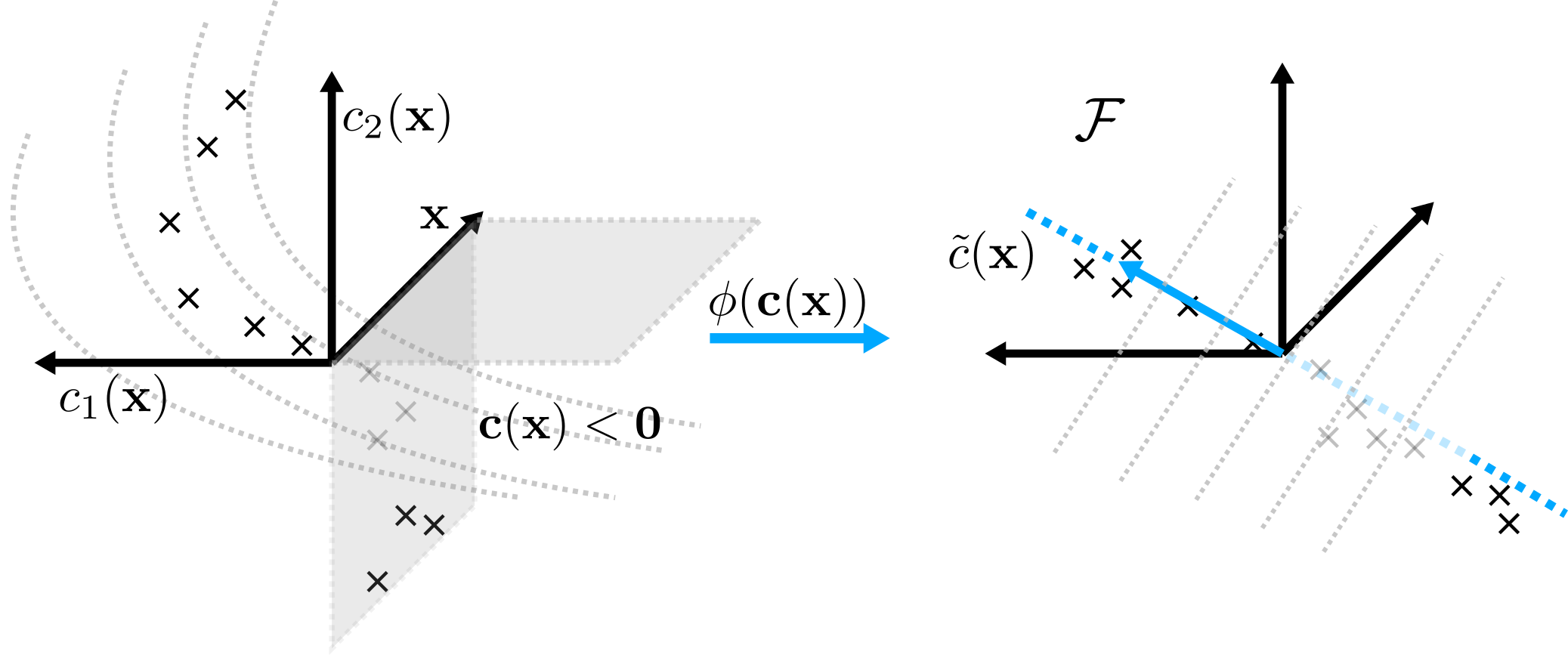}
		\caption{Kernel Principal Component Analysis}
		\label{fig:kpca_graphical}
	\end{subfigure}
	\caption{Graphical interpretation of dimensionality reduction for constraints. On the left, PCA as a linear method is depicted, finding the lower dimensional subspace (blue arrow). On the right, the nonlinear extension, kPCA, is shown, first using a nonlinear kernel to map into the infinite dimensional space $\mathcal{F}$ and subsequently perform the standard PCA. The figure is inspired by \cite{scholkopf_nonlinear_1998}.}
	\label{fig:pca_kpca}
\end{figure}

When large-scale constraints are involved, the computational time as well as the needed storage scales drastically since one $\mathcal{GP}$ model has to be constructed and trained for each constraint. Therefore, describing the constraints on a latent space allows to significantly lower the computational burden. This idea is based on the work of \cite{higdon_computer_2008}, who project the simulation output onto a lower dimensional subspace where the $\mathcal{GP}$ models are constructed. Other works extended this method then by employing, among others, kPCA as well as manifold learning techniques to account for nonlinearities \citep{xing_reduced_2015, xing_manifold_2016}. However, the aforementioned authors try to approximate PDE model simulations with high-dimensional outputs, whereas, to the best of the authors' knowledge, the combination of dimensionality reduction techniques for use in high-dimensional BO with large-scale constraints for design optimisation is novel. \\~\\
The methods herein presented are capable of extracting the earlier introduced, most important principle components of available data, reducing the required amount of $\mathcal{GP}$ models to $g$ instead of $G$, with $\textbf{v}_j$ as the $j$-th orthogonal basis vector. After projecting the data onto the lower dimensional subspace by using either PCA as in equations \ref{eq:PCAprojection1} or kPCA in equation \ref{eq:KPCAprojection}, $\mathcal{GP}$s are constructed on the latent output space as independent batch $\mathcal{GP}$s , formulated as  
\begin{equation}
	\tilde{\textbf{c}}_i \sim \hat{\tilde{\textbf{c}}}_{i}  = \mathcal{GP} \left( m_i (\textbf{x}) ,  k_i(\textbf{x}, \textbf{x}') \right) \forall i \in \{1,...,g\}.
\end{equation}
These constraint surrogates on the latent space are then used to navigate through the design space to ultimately find a feasible and optimal design. A graphical interpretation is depicted in Figure \ref{fig:pca_kpca}.
In the following, the projection of the constraints onto the lower-dimensional subspace in the $i$-th iteration is denoted as $\mathcal{P}_i: \mathbb{R}^{G} \to \mathbb{R}^{g}$.


\begin{algorithm}
	\caption{SCBO with Latent Gaussian Processes}\label{alg:scbo_rb_dynamic}
	\begin{algorithmic}
		\State \textbf{Input:} Input space $\mathcal{X}$, Number of candidates $N_c$, batch size $q_c$, number of initial samples $N_i$, SCBO hyperparameters, number of eigenvalues $N_{ev}$ or tolerance $\tau_{ev}$
		\State Compute DoE $\mathcal{D}_0 = \{ \textbf{x}_{i}, f(\textbf{x}_{i}), \textbf{c}(\textbf{x}_{i}) \}_{i=1:N_i}$
		\State $k = 0$
		\While{Computational budget is not exhausted} 
		\State With $\textbf{c}(\textbf{x}) \subset \mathcal{D}_k$ compute projection $\mathcal{P}_k$
		\State Project constraints onto lower dimensional subspace $\tilde{\textbf{c}}(\textbf{x}) = \mathcal{P}_k(\textbf{c}(\textbf{x}))$
		\State Fit $\mathcal{GP}$ for $f(\textbf{x}) \sim \hat{f}(\textbf{x})$ and $\tilde{\textbf{c}}(\textbf{x}) \sim \hat{\tilde{\textbf{c}}}(\textbf{x})$ 
		\State $\textbf{x}_+ \leftarrow$ \textsc{constrainedThomsponSampling} (see Algorithm \ref{alg:ts})
		\State Evaluate $\textbf{x}_+$ and observe $f(\textbf{x}_+), \ \textbf{c}(\textbf{x}_+)$
		\State Update \textsc{TuRBO} state
		\State $\mathcal{D}_{k+1} = \mathcal{D}_{k} \cup \{ \textbf{x}_{+}, f(\textbf{x}_+), \textbf{c}(\textbf{x}_+) \}$
		\State $k \leftarrow k+1$
		\EndWhile
	\end{algorithmic}
\end{algorithm}

\begin{figure}[h]
	\centering
	\includegraphics[width=0.65\textwidth]{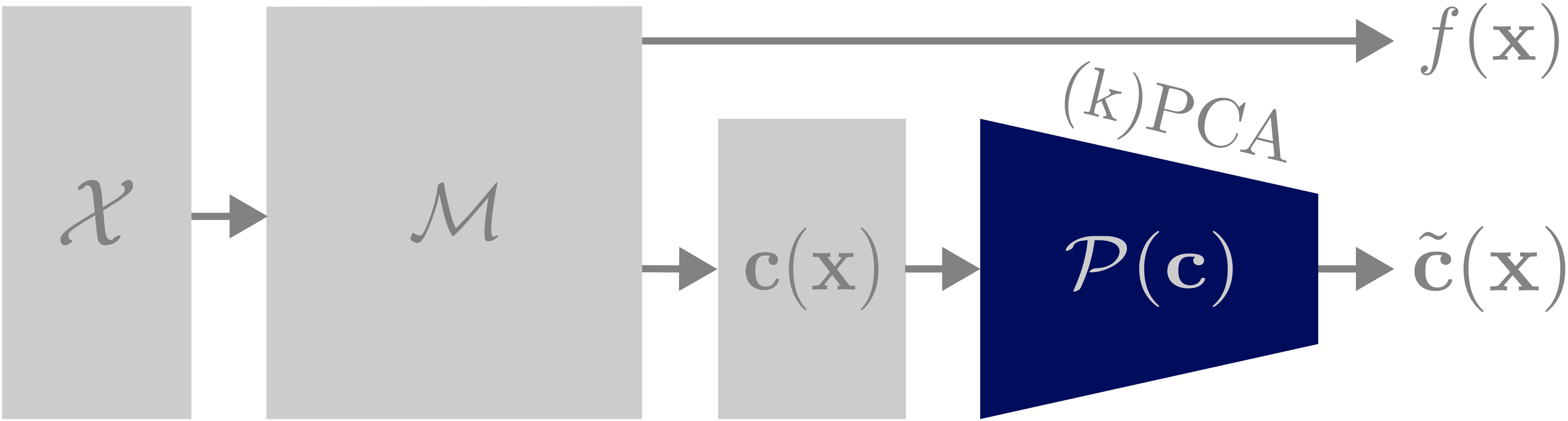}
	\caption{Schematic illustration of (k)PCA-GP: $\mathcal{M}$ denotes the numerical model, mapping from the design space $\mathcal{X}$ to the objective $f$ and constraints $\mathbf{c}$ as outputs. The constraints are then mapped via (k)PCA onto a lower-dimensional representation $\tilde{\mathbf{c}}$ where the independent $\mathcal{GP}s$ are constructed.}
	\label{fig:schematic}
\end{figure}
A schematic illustration of the \(\mathcal{GP}\) construction is presented in Figure \ref{fig:schematic}. It is important to emphasise that the validity of a feasible design, where no constraints are violated, is checked in the original space rather than within the lower-dimensional subspace. This is made possible since in each iteration, a batch of \( q \) new samples is obtained and evaluated using the expensive-to-evaluate model. Hereinafter, the two methods are called PCA-GP SCBO and kPCA-GP SCBO. 

\subsection{Related Work and Complexity Considerations}
To tackle the issue of many outputs, several works have been published. The Intrinsic Co-regionalisation Model (ICM) can be related to the Linear Model of Co-regionalisation (LMC), presented in \cite{alvarez_kernels_2012} and based on Multi-Task Gaussian Processes \cite{bonilla_2007}. However, due to taking into account inter-task correlation, the size of the covariance matrix increases drastically. While in independent \(\mathcal{GP}\) models, inference and learning typically has a complexity of \(\mathcal{O}((G+1)N^3)\) and \(\mathcal{O}((G+1)N^2)\) for storage, the size of multi-task models extends due to their Kronecker structure to complexities of \(\mathcal{O}(N^3 (G+1)^3)\) for inference and learning, with \(G+1\) denoting the number of constraints plus the objective. Similarly, the storage complexity also scales to \(\mathcal{O}(N^2 (G+1)^2)\), posing significant computational challenges when the number of tasks/constraints and/or data points becomes large.\\
The benefit of (k)PCA-GPs now is the fact that by mapping the outputs/constraints onto a $g$-dimensional subspace while no inter-task correlations are respected, the computational costs for inference and learning only scale linearly to \(\mathcal{O}((g+1)N^3 + G^3)\) where $\mathcal{O}(G^3)$ accounts for the eigendecomposition during (k)PCA and $\mathcal{O}((g+1)N^2)$ for storage, where $g \ll G$.\\
To adress some of the issues, apart from \cite{higdon_computer_2008}, 
\cite{zhe19} present scalable High-Order \(\mathcal{GP}\)s (HOGP) and show that their method is superior to (k)PCA-GP in terms of accuracy. Since (k)PCA-GPs assume a linear structure of the outputs, meaning that the output is a linear combination of bases vectors, HOGP does not impose this kind of structure, thus claiming to be more flexible. The authors in \cite{maddox_bayesian_2021} then extend Multi-Task GPs and later HOGP for a large number of outputs by employing Mathoron's rule to alleviate the computational burden of sampling from the posterior.  Additionally, \citet{bruinsma_scalable_2019} introduce a method that tackles the problem of needing a high number of linear basis vectors in PCA-GP, which still scale cubically in the dimensionality of the subspace when inter-task correlations are taken into account. They leverage the statistics of data to achieve linear scaling. However, all these works take into account inter-task correlation and thus scale poorly compared to k(PCA)-GP, as concluded by \cite{zhe19}. Due to the fact that in engineering design problems the dimensionality and constraints can become very large, thus high values for $N$ and $G$ can be expected, this work uses batched, independent \(\mathcal{GP}\)s in the reduced latent space as originally proposed by \cite{higdon_computer_2008}. Due to the use in Bayesian Optimization (BO) and the continuous retraining of the surrogates, the approach employed here significantly accelerates computations while maintaining acceptable accuracy as presented in \cite{zhe19}.

\section{Numerical Experiments}\label{ch:application}
In this section, the presented methodology is applied to a benchmark case before results for the aeroelastic tailoring optimisation problem are shown. For comparison purposes, we adopt the reasoning of \cite{hernandez-lobato_general_2016}, where a feasible solution is always preferred over an infeasible one. Therefore, we use the maximum value from all feasible solutions as the default for all infeasible solutions. To leverage the capabilities of existing, well performing frameworks, this study employs \textsc{BoTorch} \citep{balandat2020botorch} and \textsc{GPyTorch} \citep{gardner2018gpytorch} to make use of their extensive capabilities.

\subsection{$\mathbf{7D}$ Speed Reducer Problem with $\mathbf{11}$ Black-Box Constraints}\label{ch:speedreducer}
The $7D$ speed reducer problem from \cite{lemonge_2010} includes $11$ black-box constraints.

\begin{figure}[h]
	\centering
	\includegraphics[width=0.7\textwidth]{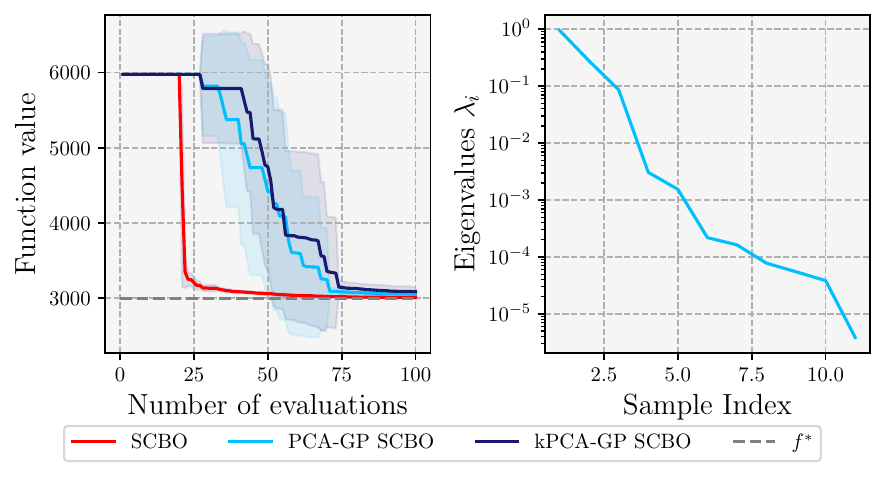}
	\caption{(\textbf{left}) $7D$ Speed reducer problem with $11$ black-box constraints from \cite{lemonge_2010}. $20$ experiments are performed, where the solid line represents the mean objective value over the $20$ experiments and the shaded area the standard deviation. $f^*$ denotes the known optimal value of this problem. (\textbf{right}) The eigenvalues of the matrix $\textbf{C}$ with $N=10$ samples are plotted}.
    \label{fig:speedreducer}
\end{figure}

The known optimal value for this problem is $f^* = 2996.3482$. The results for all three evaluated methods (SCBO, PCA-GP SCBO and kPCA-GP SCBO) are shown in Figure \ref{fig:speedreducer}. Additionally, the decay of the eigenvalues $\lambda$ of the constraint matrix $\textbf{C} \subset \mathcal{D}$ is depicted. In this example where $G=11$, $g=4$ principal components are chosen. The SCBO hyperparameters are defined according to \cite{eriksson_scalable_2021}. The batch size is defined as $q=1$ and $N=20$ initial samples. 

\begin{figure}[h]
	\centering
	\includegraphics[width=0.35\textwidth]{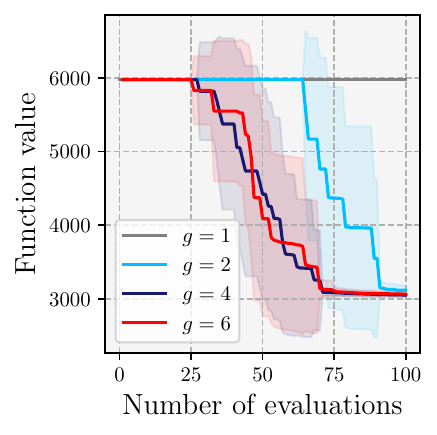}
	\caption{The influence of the number of principal components $g$ on the result.}
    \label{fig:eigval_study}
\end{figure}

The results are compared in Table \ref{table:time_speedreducer}
\begin{table}[!h]
	\caption{Computational time for speed reducer benchmark}
	\centering
	\begin{tabular}{c c c c c c}
		\textbf{Method} & \textbf{$\tilde{f}^{*}$ [-]} & \textbf{$(\tilde{f}^{*}-f^*)/f^*$[\%]} & \textbf{Time [s]} & \textbf{Time Saving [\%]} & \textbf{Successful runs [-]} \\
		\hline
		SCBO & 3007.20 & 0.36 & 501.38 & - & 20/20\\
		PCA-GP SCBO & 3053.30&1.90 & 201.38 & 59.83 & 20/20 \\
		kPCA-GP SCBO & 3088.39&3.07& 216.96 & 56.73 & 20/20\\
		\hline
	\end{tabular}
    \label{table:time_speedreducer}
\end{table}
All methods find a feasible and optimal design. It is obvious that the original SCBO method converges faster than the once employing latent $\mathcal{GP}$s. In SCBO, each constraint is modelled independently via batched $\mathcal{GP}$s. However, besides the fact that the proposed methods are significantly faster, see Table \ref{table:time_speedreducer}, it is shown that both ultimately converging to a optimum very close to the one obtained via SCBO and the analytical solution $f^*$. \\~\\
kPCA-GP SCBO uses the Gaussian kernel, written as 
\begin{equation}\label{eq:gaussian_kernel}
	k(\textbf{x},\textbf{x}') = \exp \left( \frac{-||\textbf{x}-\textbf{x}'||^2}{2\sigma^2} \right).
\end{equation}
Here, PCA-GP SCBO converges slightly faster than kPCA-GP SCBO. It needs to be emphasised that this problem also does not show a fast decay of the eigenvalues, as can be seen in Figure \ref{fig:speedreducer} (left). \\~\\
In addition, the influence of the number of principal components, $g$, is studied in Figure \ref{fig:eigval_study}. It can be observed that $g$ affects the convergence of the optimisation. Notably, when $g=2$, although convergence is slower, the mean value found is close to the analytic value $f^*$. However, when $g=1$ the subspace does not cover enough of the feasible design space, resulting in no feasible value being found.
Summarising, the lower dimensional subspace is constructed based on the constraint values in $\mathcal{D}$. Assuming that the global optimum lies on the boundary of the feasible space $\mathcal{X}_f$, the success of the method highly depends on how accurately the lower dimensional subspace captures the original space. That stresses the importance of computing the projection matrix $\mathcal{P}_i$ in every iteration. However, we find that for this specific case fixing $\mathcal{P}_i = \mathcal{P}_0$, the Algorithm \ref{alg:scbo_rb_dynamic} exhibits a better performance, presumably due to the rather low dimensionality and low number of constraints in combination with the use of the Trust Region. 

\subsection{Aeroelastic Tailoring: An MDO Problem with $\mathbf{108D}$ and $\mathbf{1786}$ Black-Box Constraints}
The MDO problem of aeroelastic tailoring addressed in this work presents a high-dimensional problem with large-scale constraints, involving both high-dimensional inputs and outputs. Unlike the aforementioned benchmark problem where it is practical to construct a $\mathcal{GP}$ for each constraint, this is computationally infeasible here, where the number of constraints is $10^{3} < G < 10^{5}$. Therefore, the methodology presented in this study facilitates the process by modelling these constraint $\mathcal{GP}$s in a latent space.

\begin{figure}[h]
	\centering
	\includegraphics[scale=0.65]{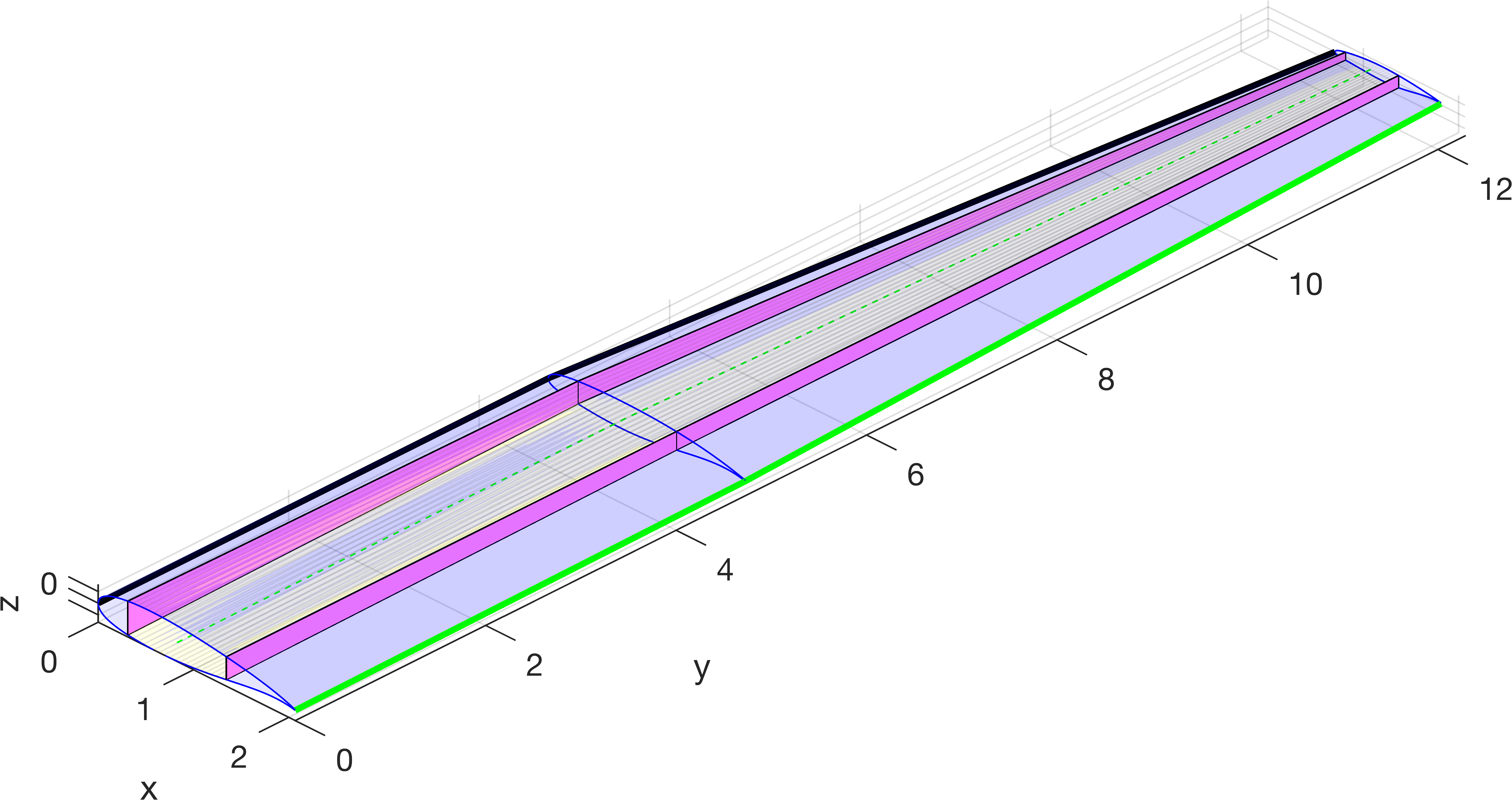}
	\caption{Wing structure consisting of wingbox and airfoil shape. The proposed problem optimises the stiffness and thickness of the wingbox. The wingbox is span-wise discretised in three sections, where top skin, bottom skin, front spar and rear spar can take on different stiffness and thickness values. The wing span exhibits $b=12.28$m, with a $c=2.068$m chord at the root and $c=1.113$m chord at the tip. The front and rear spar are located at $0.15c$ and $0.65c$, respectively. }
	\label{fig:wing}
\end{figure}

Figure \ref{fig:wing} depicts the wing to be aeroelastically tailored. More precisely, the structural tailoring is performed on the wingbox region, where the stiffness and thickness of the top skin, bottom skin, front spar and rear spar can be optimised. Additionally, the wing is discretised span-wise in three design sections. In total, $D=108$ design variables are defined, consisting of the lamination parameters $\xi \in [-1,1]$ and the thickness $t \in [0.002,0.03]$ of each panel, respectively. Each panel is described by a set of parameters $\textbf{x}^{lam}_i$. 
\begin{equation}\label{eq:desvars}
\begin{aligned}
	\textbf{x} = \biggl\{ \textbf{x}^{lam}_1, \textbf{x}^{lam}_2, ...,  \textbf{x}^{lam}_{n_p} \biggr\} \in \mathbb{R}^{108} && \text{with} && \textbf{x}^{lam}_i = \biggl\{ \xi_1^A, \xi_2^A, \xi_3^A, \xi_4^A, \xi_1^D, \xi_2^D, \xi_3^D, \xi_4^D, t \biggr\} \in \mathbb{R}^9.
\end{aligned}
\end{equation}
Based on the classical laminate theory, the following constitutive equations are used to relate the distributed forces $N$ and moments $M$, with the in-plane $\epsilon^0$ and curvature $\kappa$ strains
\begin{equation}
\begin{bmatrix} N \\ M \end{bmatrix} =\begin{bmatrix} \textbf{A}(\textbf{x}) & \textbf{0} \\ \textbf{0} & \textbf{D}(\textbf{x}) \end{bmatrix} \begin{bmatrix} \epsilon^0 \\ \kappa \end{bmatrix}
\end{equation}
The so-called ABD-matrix can be calculated by means of lamination parameters according to \cite{tsai-pagano-1968} as follows:\\
\begin{equation}\label{eq:lam_param}
\begin{aligned}
    \textbf{A(x)} = t(\pmb{\Gamma_0} 
    + \pmb{\Gamma_1}\xi_1^A
    + \pmb{\Gamma_2}\xi_2^A
    + \pmb{\Gamma_3}\xi_3^A
    + \pmb{\Gamma_4}\xi_4^A
    )\\
    \textbf{D(x)} = \frac{t^3}{12}(\pmb{\Gamma_0} 
    + \pmb{\Gamma_1}\xi_1^D
    + \pmb{\Gamma_2}\xi_2^D
    + \pmb{\Gamma_3}\xi_3^D
    + \pmb{\Gamma_4}\xi_4^D
    )\\
\end{aligned}
\end{equation}
\noindent where $\pmb{\Gamma_i}$ are material invariants, defined in \cite{tsai-pagano-1968}. Equation \ref{eq:lam_param} encodes the dependency of the design variables $\textbf{x}$ with the stiffness of the system \citep{daniel_engineering_2006}. The constraints result from the incorporation of two loadcases. These multiple loadcases are often one of the reason why the number of constraints can become very high. The aforementioned constraints arise from the multidisciplinary analyses, summarised in Table \ref{table:constraints} and leading to a total number of $G=1786$, similarly depending on the input variables $\textbf{x}$. More information of the aeroelastic tailoring optimisation problem can be found in \cite{maathuis_high-dimensional_2024}.
\begin{table}[!h]
	\caption{Aeroelastic tailoring constrained optimisation problem}
	\centering
	\begin{tabular}{c c c c}
		\hline\hline
		\textbf{Type} & \textbf{Parameter} & \textbf{Symbol} &\textbf{/Loadcase}\\
		\hline
		Objective & Minimise Wing Mass & $f$ & \\
		\hline
		\multirow{2}{*}{Design Variables ($D$)} & Lamination Parameter & \multirow{2}{*}{$\textbf{x}$} & \\
		& Laminate Thickness & & \\
		\hline
		\multirow{6}{*}{Constraints ($G$)} & Laminate Feasibility & $\boldsymbol{c}_{lf}$ & No \\
		& Static Strength & $\boldsymbol{c}_{tw}$ & Yes\\
		&  Buckling & $\boldsymbol{c}_{b}$ & Yes \\
		& Aeroelastic Stability & $\boldsymbol{c}_{ds}$ & Yes \\
		& Aileron Effectiveness & $c_{ae}$ & Yes\\
		& Local Angle of Attack & $\boldsymbol{c}_{AoA}$ & Yes \\
		\hline\hline
	\end{tabular}
	\label{table:constraints}
\end{table}
Apart from the mathematical reasoning to find a latent space of the output data, the premise of the introduced methodology lies in the consistency of the physics governing the constraints across loadcases, where eventually only the load changes. This stresses the potential for compressing this information due to the unchanged underlying physics for varying loadcases.\\~\\
The lamination parameter feasibility constraints are, however, closed-form equations. These analytical equations do not need to be modelled via surrogates since their behaviour is known in the design space. Thus, these constraints are taken into account inherently within the sampling process via rejection sampling. Every candidate point in $N_c$ is only added if not violating one of these feasibility constraints.

\begin{figure}[h]
	\centering
    \includegraphics[width=0.7\textwidth]{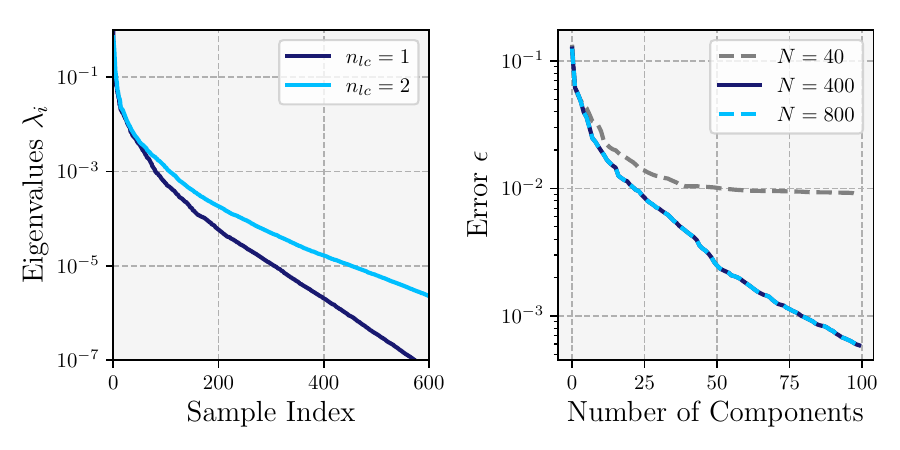}
	\caption{Investigating the constraints in $\mathcal{D}$. (\textbf{left}) Showing the decay of the eigenvalues for $N=416$, performing PCA on the matrix $\textbf{C}$ for one and two loadcases. (\textbf{right}) Computing the error for PCA depending on how many principal components are taken into account (Equation \ref{eq:error}).}
    \label{fig:at_eigs}
\end{figure}

\begin{figure}[h]
	\centering
    \includegraphics[width=0.8\textwidth]{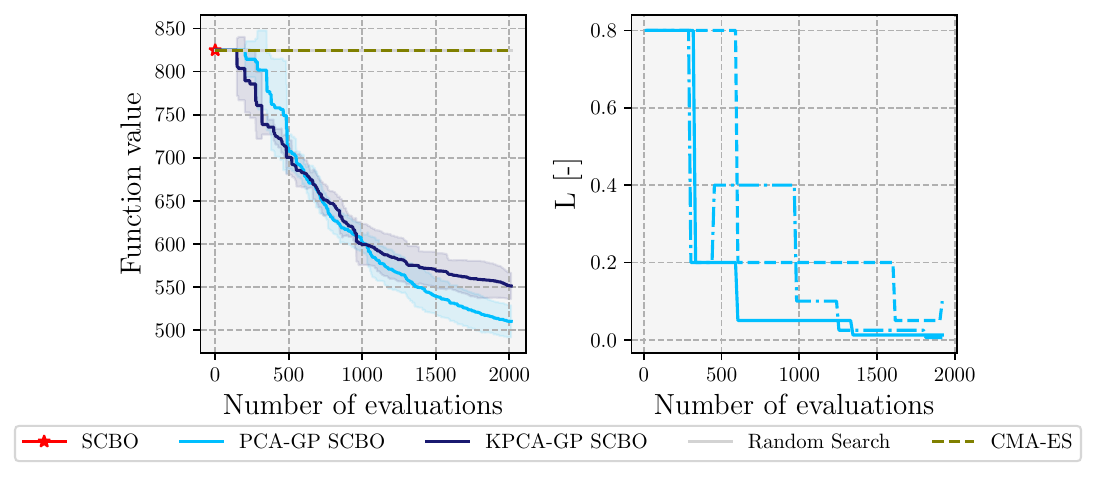}
	\caption{(\textbf{left}) Optimisation results of aeroelastic tailoring with $D=108$, $G=1786$, $g=35$ and $N = D$ (\textbf{right}) History of Trust Region hyper-rectangle size $L$ (Equation \ref{eq:TRlength}).}
    \label{fig:atresults}
\end{figure}

The aforementioned aeroelastic tailoring model is used to compute the DoE $\mathcal{D}$ with $N=416$ samples. Sampling was performed via Latin Hypercube Sampling (LHS). Subsequently, PCA is applied on the matrix $\textbf{C}$ to investigate its eigenvalues. Figure \ref{fig:at_eigs} depicts the decay of these computed eigenvalues. If the same error metric as in Subsection \ref{ch:speedreducer}, eigenvalues up to approx $\lambda_i \approx 10^{-2}$, thus $g = 29$ principal components might be enough to construct a lower dimensional subspace of sufficient accuracy. \\~\\
As previously noted, the high number of constraints stems from the incorporation of multiple loadcases. Consequently, it becomes intriguing to explore how the eigenvalues vary when the number of loadcases is altered. Recall that the eigenvalues denote the importance of their corresponding eigenvector, which serves as a measure of where to truncate the projection matrix. Beyond that, in Figure \ref{fig:at_eigs} we compared the eigenvalues of $n_{lc} = 1$ and $n_{lc} = 2$ loadcases. It can be observed that, even though the number of constraints in the original space has doubled, from $G=893$ to $G=1786$, if the eigenvalues $\lambda_i > 10^{-2}$ are used, no more principal components have to be taken into account. For $\lambda_i > 10^{-3}$, however, only $27$ more components are needed to maintain the same error. Beyond that, the threshold of the eigenvalues is commonly set based on experience, thus can be seen as a hyper-parameter of the method. \\
To compute the projection error, some unseen data $\textbf{C}_{*}$, is mapped onto the lower dimensional subspace $\tilde{\textbf{C}}_* = \boldsymbol{\Psi}_g^{\top} \textbf{C}_{*}$. Since PCA is a linear mapping, the inverse mapping can be simply computed by  $\hat{\textbf{C}}_{*} = \tilde{\textbf{C}}_* \boldsymbol{\Psi}$. The approximation error can then be computed by
\begin{equation}\label{eq:error}
	\epsilon = \frac{ \parallel \textbf{C}_* - \hat{\textbf{C}}_* \parallel_{F}^{2} }{ \parallel \textbf{C}_* \parallel_{F}^{2}}.
\end{equation}
In Figure \ref{fig:at_eigs} (right), the trend reveals that including more components leads to a reduced error, even for unseen data. Furthermore, to investigate how the construction of the lower-dimensional subspace behaves with sample size variation, the error $\epsilon$ is shown for $N=40$, $N=416$ and $2N$ samples. It can be seen that the error is approximately the same for the latter two cases. As anticipated, an insufficient initial sample size $N$ results in limited information availability during the subspace construction, consequently leading to a larger error. Moreover, the conclusion drawn is that even with $N=416$ samples, sufficient data is available to attain a reasonable subspace. Furthermore, increasing the number of samples in the DoE does not contribute to higher accuracy. To mitigate this issue, the projection matrix is recalculated in every iteration of the optimisation process to incorporate as much data as possible.\\~\\
Figure \ref{fig:atresults} (left) shows the results for the $108D$ aeroelastic tailoring problem, comparing the results of SCBO, (k)PCA-GP SCBO, Random Search and CMA-ES \citep{hansen2006}. Again, kPCA-GP SCBO uses the Gaussian kernel defined in Equation \ref{eq:gaussian_kernel}. A total of 5 experiments are performed per method on a conventional computer with: \textsc{Intel Xeon w3-2423, 6 cores, 32gb RAM}. The original SCBO method crashes due to insufficient memory after the first iteration, while trying to construct $1786$ high-dimensional $\mathcal{GP}$ surrogates. However, a good convergence can be observed for the PCA-GP SCBO and kPCA-GP SCBO, with $g=35$, where again PCA performs better than kPCA. Additionally, it is important to note that given the size of the DoE $\mathcal{D}_0$  being \( N = D \), a feasible design point can be efficiently identified, even if all points in the DOE at iteration \( k = 0 \) were initially infeasible. This is also highlighted by the results of the random search and CMA-ES which both fail in finding a feasible point.
Due to the high-dimensional design space and the high number of constraints, the probability of finding a feasible point where no constraints are violated is extremely low with random search, which was unable to find a single feasible design point. Therefore, the proposed method renders the observed advantage of finding efficiently feasible points even when $\mathcal{D}_0$ only contains infeasible ones. Figure \ref{fig:atresults} (right) illustrates the size of the trust region over the number of model evaluations for three randomly chosen runs. It can be observed that the size generally decreases. However, as seen for instance in the dark blue curve, the optimiser occasionally gets stuck, increases the trust region size to escape the locality while the evaluation budget is not exhausted, and then restarts to decrease it.

\begin{figure}[h]
	\centering
	\includegraphics[width=0.5\textwidth]{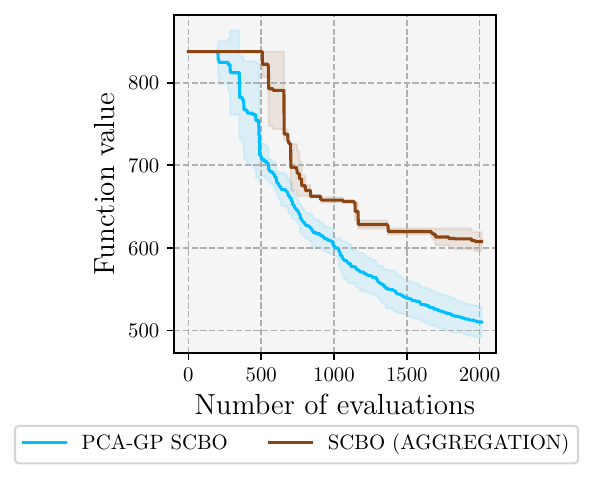}
	\caption{Comparison of best result with constraint aggregation}
	\label{fig:aggragetion}
\end{figure}

For the sake of completeness, we compare the proposed method to the so-called constraint aggregation approach, using the Kreisselmeier-Steinhauser (KS) function, written as 
\begin{equation}\label{eq:ks-function}
	KS (\textbf{x}) = c_{max} + \frac{1}{\rho} \log \left[ \sum_{j=1}^{m} e^{\rho c_j(\textbf{x})} \right].
\end{equation}
This function aggregates multiple constraints, arising for example from a buckling or strength analysis into one constraint function. We implement this to lower the number of needed surrogates and compare the results against the best candidate so far. We aggregate the strain and buckling constraints for each loadcase individually for which we construct the $\mathcal{GP}$, while the other constraints are modelled independently, leading to a reduced reduced number of constraints $g = 66$. It should be noted that, compared to PCA-GP SCBO/ kPCA-GP SCBO where $g=35$ principal components were used, in the aggregation approach 66 surrogate models need to be constructed, needing approximately twice as long for surrogate construction. Thus, downsides are the increased number of needed surrogate models in high-dimensional space as well as the additional hyperparameters needed to define which constraints to aggregate as well as the hyperparameter $\rho$ which we set in this case to $\rho = 100$. For more information the reader is referred to \cite{Martins2005d}. \\~\\
It should be pointed out that in the constraint aggregation case not only requires more $\mathcal{GP}$s to be constructed, increasing the need for computational resources but also the structure of the constraints need to be known such that only constraints arising from one discipline are aggregated. This is additionally needed knowledge which might be not available, drastically lowering the generality of this approach. The corresponding results can be found in Figure \ref{fig:aggragetion} where we compare SCBO with the aggregation technique with PCA-GP SCBO.

Hypothesising why the aggregation method performed worse than the herein introduced approaches is first of all its conservativeness and second, the high-order of the output function due to approximating all the constraints, possibly leading to a quasi-non-smooth function which is cumbersome to approximate. However, further research has to be performed to confirm these statements. 

\section{Conclusion and Future Work}
The aeroelastic tailoring problem exemplifies a high-dimensional multidisciplinary design optimisation challenge characterised by large-scale constraints. Conducting a global design space search is inherently complex, particularly when dealing with black-box optimisation problems where computing gradients is problematic. Constrained BO faces scalability issues due to the extensive number of constraints involved. To mitigate the scalability shortcomings of the aforementioned methods, $\mathcal{GP}$s are constructed on the latent space of the high-dimensional outputs in combination with Trust Region-based approach. By significantly reducing the number of required $\mathcal{GP}$s, substantial computational savings can be realised, making certain problems feasible and aligning with the objectives to reduce computational expenses. These savings are even more pronounced in high-dimensional settings where the training of each $\mathcal{GP}$ is critical.\\~\\
Within aeroelastic tailoring, feasible designs can be found relatively easily by increasing the thickness of each panel. However, this simplicity does not extend to other problems. The presented approach demonstrates the capability to drastically reduce computational time, thus making SCBO feasible for such problems. Numerical investigations confirm the applicability of this method to aeroelastic tailoring, showcasing its effectiveness for multiple load cases with minimal additional principal components required.\\~\\
An analytical example further illustrates that the proposed method converges to approximately the same objective function value. While our work primarily addresses aeroelastic tailoring, the method’s generality allows for application to various problems involving large-scale constraints. This flexibility is supported by numerical evidence showing the ease of application to diverse high-dimensional constraint problems.\\~\\
Additionally, any dimensionality reduction method, such as autoencoders, can be seamlessly integrated into the methodology. When compared to other methods for handling large-scale constraints, such as penalty and constraint aggregation methods, our proposed method demonstrates superior results without relying on specific knowledge about constraint categories. While the herein presented method works with a fixed user-defined or eigenvalue-based number of principal components $g$, a promising path could be an extension of this method, using an adapting number $g$ such that the approximation error of the latent space is minimised. This might further improve the method. Moreover, future research will focus on simultaneously reducing input and output spaces. Our current methodology necessitates training latent $\mathcal{GP}$s on the full-dimensional input space, limiting the scalability. Approaches like REMBO  \citep{wang_bayesian_2016} , ALEBO  \citep{letham_re-examining_2020} and (k)PCA-BO \citep{raponi_2020,antonov_2022} offer promising avenues for further reducing computational costs during hyperparameter tuning. Simultaneously reducing input and output space would highly increase the scalability of this approach. Moreover, the efficient utilisation of gradients, if available, will be explored to combine gradient-based and surrogate approaches. This could facilitate the use of active subspaces, potentially enhancing performance. \\~\\
Besides its application in BO, this research also holds promise for design under uncertainty. $\mathcal{GP}$s offer a distinct advantage in providing a measure of variance. When addressing systems with high-dimensional outputs, this method becomes particularly advantageous. By leveraging Principal Component Analysis (PCA), we facilitate an efficient mapping back to the original high-dimensional space. This approach is particularly pertinent for engineering challenges where multiple model outputs are commonplace, offering a scalable solution for variability assessment. Furthermore, our method's potential application in a multi-fidelity optimisation strategy will be explored to bolster computational efficiency and practical feasibility.

\section*{Acknowledgments}
The authors would like to express their sincere gratitude to Embraer S.A., especially Pedro Higino Cabral and Alex Pereira do Prado, for their invaluable support and collaboration within the Aeroelastic Tailoring Enabled Design project. Their expertise, resources, and guidance have been instrumental in the successful completion of this study.

\bibliographystyle{unsrtnat}
\bibliography{references.bib}  


\end{document}